\documentclass[amsmath,amssymb,eqsecnum,showpacs,nofootinbib,superscriptaddress,twocolumn]{revtex4-1}
\usepackage{chngcntr}
\usepackage{graphicx}
\usepackage{bm}
\usepackage{dcolumn}
\usepackage{color}
\usepackage{caption, subcaption}
\usepackage{amssymb,amsmath, mathrsfs}
\usepackage{amssymb, graphics, setspace}
\usepackage{hyperref}
\usepackage[all]{xy}

\newcommand{\rc}[1]{}

\newcommand{\be}{\begin{equation}}
\newcommand{\ee}{\end{equation}}
\newcommand{\bea}{\begin{eqnarray}}
\newcommand{\eea}{\end{eqnarray}}

\def\sss{\scriptscriptstyle}

\begin{document}

\counterwithout{equation}{section}

\author{Roberto~Balbinot}
\affiliation{Dipartimento di Fisica e Astronomia dell'Universit\`a di Bologna
  and INFN sezione di Bologna, Via Irnerio 46, 40126 Bologna, Italy}

\author{Alessandro~Fabbri}
\affiliation{Departamento de F\'isica Te\'orica and IFIC,
  Universidad de Valencia-CSIC, Calle Dr. Moliner 50, 46100 Burjassot, Spain}

\author{Giorgio~Ciliberto}
\affiliation{Laboratoire d'Informatique de Paris 6, CNRS,
  Sorbonne Universit\'e, 4 place Jussieu, 75005 Paris,
  France}
\affiliation{Universit\'e Paris-Saclay, CNRS, LPTMS, 91405
  Orsay, France}
\affiliation{Physikalisches Institut, Albert-Ludwigs-Universit\"at Freiburg,
  Hermann-Herder-Stra$\beta e$ 3, 79104 Freiburg, Germany}

\author{Nicolas~Pavloff}
\affiliation{Universit\'e Paris-Saclay, CNRS, LPTMS, 91405 Orsay, France}
\affiliation{Institut Universitaire de France}

\title{Backreaction equations for 1+1 dimensional BEC sonic black holes}

\begin{abstract} 
As in the gravitational context, one of the most challenging open
questions in analogue black holes formed in Bose-Einstein condensates
concerns the backreaction of Hawking-like radiation on the condensate
and its subsequent evolution. In this work we derive the basic
equations describing this backreaction within the density-phase
formalism, which avoids infrared divergences and is particularly well
suited to one-dimensional configurations.
\end{abstract}

\date{\today}

\maketitle

In 1974 Hawking \cite{hawking74, hawking75} showed that, taking into
account Quantum Mechanics, black holes (BHs) are not at all ``black"
as believed but emit thermal radiation and hence ``evaporate". The
evolution of a BH driven by the emission of quantum fields is commonly
referred to as the ``backreaction".

In the absence of a complete and self consistent quantum gravity
theory, this backreaction is described within the framework of Quantum
Field Theory in curved spacetime by the ``semiclassical" Einstein
equations \cite{birrell-davies}
\begin{equation} \label{uno}
  R_{\mu\nu}-\frac{1}{2}Rg_{\mu\nu}=
  8\pi G \langle T_{\mu\nu}(\hat\phi,g_{\mu\nu})\rangle\ ,
\end{equation}
where on the l.h.s. we have the Einstein tensor while the r.h.s.  is
the expectation values in the Unruh state \cite{unruh76} of the stress
energy tensor operator for the quantum fields $\hat\phi$ (matter and
gauge fields) evaluated on the spacetime metric $g_{\mu\nu}$ which is
treated as a classical (not quantum) field. These equations, together
with the field equation $F(\hat\phi,g_{\nu\nu})=0$ for the quantum
fields $\hat\phi$, have to be solved self consistently for the metric
$g_{\mu\nu}$ which would then describe the spacetime of the
evaporating BH.

One expects this scheme to give a truthful description of the process
valid on scales much bigger than the Planck scale $(\sim
10^{-33}$ cm) where quantum effects associated to the gravitational
field can no longer be neglected. Unfortunately, even in the simplest
case of spherical symmetry one does not know $ \langle
T_{\mu\nu}(\hat\phi,g_{\mu\nu})\rangle$ for an arbitrary spherically
symmetric BH metric. Some insight into the backreaction has come by
considering a perturbative approach (\`a la Hartree-Fock) to
Eq. (\ref{uno}). One considers first the classical Einstein vacuum
equations $R_{\mu\nu}=0$ (i.e., Eq. (\ref{uno}) with no quantum source
on the r.h.s.) leading to a stationary BH solution
$g_{\mu\nu}^{0}$. In spherical symmetry $g_{\mu\nu}^0$ would be the
metric of a Schwarzschild BH. Then one sets
$g_{\mu\nu}=g_{\mu\nu}^0+\delta g_{\mu\nu}$, where $\delta g_{\mu\nu}$
represents the (supposed) small semiclassical correction to the
background $g_{\mu\nu}^0$, and rewrite the backreaction
eqs. (\ref{uno}) as
\begin{equation} \label{due}
  R_{\mu\nu}-\frac{1}{2}Rg_{\mu\nu}=
  8\pi G \langle T_{\mu\nu}(\hat\phi,g_{\mu\nu}^0)\rangle\ ,
\end{equation}
where the l.h.s. has to be linearized in $\delta g_{\mu\nu}$ and on
the r.h.s. we have the expectation value of the energy-momentum tensor
evaluated on the background $g_{\mu\nu}^0$ using for $\hat\phi$ the
quantum field equation $F(\hat\phi, g_{\mu\nu}^0)=0$.  This should
give us a hint on how the backreaction of the quantum fields starts to
destabilise the classical solution.

From the information we have on $\langle
T_{\mu\nu}(\hat\phi,g_{\mu\nu}^0)\rangle$ near the horizon for the
Schwarzschild metric one can show that, as expected, as a consequence
of the quantum emission the BH mass decreases and the horizon shrinks
\cite{york, bardeen, balbinot}. This shrinking is caused by the
absorption by the hole of negative energy-density from the vacuum
polarization surrounding the horizon.  This is the backreaction in the
near horizon region \cite{birrell-davies}.

To overcome the problem of the lack of knowledge of the quantum source
term in the backreaction eqs. (\ref{uno}), many efforts have been
devoted to the study of two-dimensional (one space and one time
dimension) BH models. In 2D, restricting ourselves to conformal
invariant quantum fields, the expectation value $\langle
T_{\mu\nu}(\hat\phi,g_{\mu\nu})\rangle$ for an arbitrary 2D spacetime
metric is known \cite{dfu}. However in 2D the l.h.s. of
Eqs. (\ref{uno}) (the Einstein tensor) vanishes identically. So one
has to modify the original gravitational theory (General Relativity)
by introducing extra classical fields (dilatons) besides the metric to
obtain a nontrivial dynamics \cite{cghs, rst, libro-fn}. In these
models one can solve the modified backreaction equations self
consistently for the evaporating BH metric. However it is not clear
how these 2D results are significant for the real 4D world.

It is by now well established that Hawking BH radiation is not
peculiar to gravitational physics. As shown by Unruh \cite{unruh81},
phonons in a fluid with a transonic flow experience an effective BH
metric leading to an acoustic analogue of Hawking radiation. This
radiation has been (indirectly) detected in a series of experiments
performed by J. Steinhauer \cite{jeff2016, jeff2019, jeff2021} and his
group using Bose-Einstein condensates (BECs).

Following this, the most natural question which comes to mind is how
these BEC acoustic BHs evolve taking into account Hawking radiation,
i.e. the backreaction \cite{bffp, bff, suxf, ltt, brf}. Although we
have some experimental insight on how this backreaction proceeds
\cite{jeff2021}, we still lack a theoretical description of this
process. The starting point seems promising: unlike gravity we have in
this case a solid underlying quantum theory described by the
Heisenberg equation for the fundamental Bose quantum operator $\hat
\Psi(\vec x,t)$ \cite{ps}
\begin{equation} \label{tre}
  i\hbar \frac{\partial \hat \Psi}{\partial t}=-\frac{\hbar^2}{2m}
  \vec\nabla^2\hat\Psi +\left[U-\mu+g\hat\Psi^+\hat \Psi \right] \hat \Psi,
\end{equation}
which obeys the standard Bose equal time commutation rules 
\begin{eqnarray}
  &&\left[ \hat \Psi(\vec x,t),
    \hat \Psi^\dagger (\vec x',t)\right]
  =\delta^3(\vec x-\vec x')\ , \label{quattroa} \\
&&\left[ \hat \Psi(\vec x,t), \hat \Psi (\vec x',t)\right] =0=\left[
    \hat \Psi^\dagger (\vec x,t), \hat \Psi^\dagger (\vec x',t)\right]
  \ .\label{quattrob}
\end{eqnarray}
In Eq. (\ref{tre}) $\mu$ is the
chemical potential, $g$ ($>0$) the interatomic coupling, $U(\vec x)$
the external potential and $m$ is the mass of the bosonic particle.

Following the Bogoliubov approach \cite{ps} one splits the field operator $\hat
\Psi$ in a classical field $\phi(\vec x,t)$ describing the condensate
and a quantum field $\hat\psi$ describing the quantum fluctuation on
top of the condensate, i.e.
\begin{equation} \label{cinque}
  \hat\Psi(\vec x,t)= \phi(\vec x,t)+\hat\psi(\vec x,t),
\end{equation}
with  $\phi=\langle \hat \Psi \rangle$, hence $\langle
  \hat\psi \rangle=0$. 
The expectation value of
Eq. (\ref{tre}) reads \cite{zaremba}
\begin{eqnarray} \label{sei}
  i\hbar\partial_t \phi &=&
  \left[-\frac{\hbar^2}{2m}\vec\nabla^2 +U-\mu
    +g|\phi |^2+2g\tilde n\right] \phi \\
  \nonumber &+& g\tilde m\phi^*
  +g\langle \hat\psi^\dagger \hat\psi\hat\psi\rangle\ ,
\end{eqnarray}
where $\tilde n=\langle \hat\psi^\dagger \hat\psi\rangle$ and $\tilde
m=\langle \hat\psi \hat \psi\rangle$ are denoted as normal and
anomalous averages respectively. Eq. (\ref{sei}) is the backreaction
equation, while the quantum field $\hat\psi$ satisfies the quantum
operator equation \cite{zaremba}
\begin{eqnarray}
  \label{sette} &&i\hbar \partial_t\hat\psi=
  \left[-\frac{\hbar^2}{2m}\vec\nabla^2 +U-\mu
    +2g|\phi |^2\right]\hat\psi
  +g\phi^2\hat\psi^\dagger \\
  &&+2g\phi(\hat\psi^\dagger\hat\psi-\tilde n)
  +g\phi^*(\hat\psi\hat\psi-\tilde m)
  +g(\hat\psi^\dagger\hat\psi\hat\psi-
  \langle \hat\psi^\dagger \hat\psi\hat\psi\rangle)\ . \nonumber
\end{eqnarray} 
So far there is no approximation: Eqs. (\ref{sei}) and (\ref{sette})
describe self consistently the reciprocal effects of the condensate
$\phi$ and the fluctuation $\hat\psi$. The expectation values entering
the backreaction equation depend on $\phi$ which is then in principle
determined self consistently solving Eq. (\ref{sei}). Unfortunately
for a three (even two) spatial dimensional condensate the explicit
form of these expectation values for a sonic black hole profile is out
of reach of present calculations.

However, unlike the gravitational setting, with a BEC it is physically
consistent to consider configurations with one spatial
dimension. These are experimentally realised by freezing out the
transverse degrees of freedom, and are the ones used by Steinhauer in
his famous experiments \cite{jeff2016, jeff2019, jeff2021}. So from
now on we restrict ourselves to the one dimensional case which should
in principle simplify computations.  However the calculations of the
expectation values entering Eq. (\ref{sei}) are much more complicated
than in gravity since in the BEC the dispersion relation is a fourth
order equation while in gravity it is second order. Not having even in
one dimension the explicit form of the expectation values, one has to
proceed perturbatively. The idea is similar to the one used in gravity
(see Eq. (\ref{due})). If one neglects the contributions of the
quantum fields in Eq. (\ref{sei}) one obtains the Gross-Pitaevskii
equation whose solution we denote as $\phi_{\sss GP}$
\begin{equation} \label{otto}
  i\hbar \frac{\partial_t\phi_{\sss GP}}{\partial t}=
  \left[-\frac{\hbar^2}{2m}\partial_x^2
    +U-\mu +g|\phi_{\sss GP}|^2\right] \phi_{\sss GP}\ .
\end{equation}
Similarly, in the quantum equation (\ref{sette}) we neglect the second
and third order terms and the expectation values and replace $\phi$ by
$\phi_{\sss GP}$ obtaining the Bogoliubov-de Gennes equation for the
quantum field $\hat\psi$ describing the fluctuations above the
condensate background $\phi_{\sss GP}$, namely
\begin{equation} \label{nove} 
  i\hbar\frac{\partial\hat\psi}{\partial t}= \left[
    -\frac{\hbar^2}{2m}\partial_x^2
    +U-\mu
    +2g|\phi_{\sss GP}|^2\right] \hat\psi +g\phi_{\sss
    GP}^2\hat\psi^\dagger\ .
\end{equation}
 Our aim is to calculate the deviation from the leading
  order classical background $\phi_{\rm\sss GP}$ induced by quantum
  fluctuations.  We set
  $\phi=\phi_{\sss GP}+\delta\phi $ and rewrite the backreaction
  equation \eqref{sei} linearizing in $\delta\phi$ and including quantum contributions up to second
  order in $\hat{\psi}$:
\begin{eqnarray}\nonumber
  i\hbar \partial_t \delta \phi &=&
  \left[
    -\frac{\hbar^2}{2m}\partial_x^2
    +U-\mu
    +2g|\phi_{\sss GP}|^2 \right]\delta\phi  +g\phi_{\sss GP}^2 \delta \phi^* \\
  \label{undici} &+& 2g\tilde n_{\sss GP}\phi_{\sss GP}
  + g\tilde m_{\sss GP}\phi_{\sss GP}^*\ , \end{eqnarray} where the
expectation values $\tilde n_{\sss GP}$ and $\tilde
  m_{\sss GP}$ are calculated on the Gross-Pitaevskii background
$\phi_{\sss GP}$ using the Bogoliubov-de Gennes Eq. (\ref{nove}) for
the quantum field $\hat\psi$.

Unfortunately this scheme does not work as we face immediately a
problem:  $\tilde n_{\sss GP}$ and $\tilde m_{\sss GP}$
are infrared divergent in one dimension, preventing us to calculate
the backreaction using Eq. (\ref{undici}).

To overcome this problem we will use for the fundamental quantum field
$\hat\Psi$ an amplitude-phase formalism proposed first by Popov
\cite{popov1, popov2}. We rewrite $\hat\Psi$ as \begin{equation}\label{dodici}
\hat\Psi= e^{i(\theta+\hat\theta)}\sqrt{\rho(1+\hat\eta)}\ , \end{equation} where
$\theta$ and $\rho$ are respectively the classical phase and density
of the condensate field and $\hat\theta$ and $\hat\eta$ the related
phase and density fluctuations.  These operators satisfy the
commutation rule \begin{equation}\label{tredici} \left[ \hat\eta(x,t),\hat
  \theta(x',t)\right]=\frac{i}{\rho}\delta(x-x')\end{equation} and
$\langle \hat\eta\rangle=0=\langle \hat\theta\rangle$.

Expanding the field $\hat\Psi$ as given in \eqref{dodici} in terms of
$\hat\theta$ and $\hat{\eta}$ and comparing the resulting expectation
value with that of \eqref{cinque} yields
\begin{equation}\label{sedici}
  \phi=\varphi \left( 1+ \tfrac{i}{2}
  \langle\hat{\theta}\hat{\eta}\rangle
  -\tfrac12\langle \hat\theta^2\rangle
  -\tfrac18 \langle \hat{\eta}^2\rangle + ..\right),
\end{equation}
where $\varphi\equiv \sqrt{\rho}e^{i\theta}$.  Expression
\eqref{sedici} shows that, whereas the classical fields $\phi$ and
$\varphi$ are identical at leading (Gross-Pitaevskii) order, they
differ at higher order. More importantly, at the order of our
treatment, $\varphi$ is a well defined quantity which obeys a regular
equation [see Eqs. \eqref{ventiduea} and \eqref{ventidueb} below]
while $\phi$ does not.

As done before we consider a small deviation from the Gross-Pitaevskii
solution $\varphi_{\sss GP}=\sqrt{\rho_{\sss GP}}e^{i\theta_{\sss
    GP}}$, i.e. $\varphi=\varphi_{\sss GP}+\delta\varphi$. Setting for
the density and phase backreaction corrections $\rho=\rho_{\sss
  GP}+\delta\rho\ , \theta = \theta_{\sss GP}+\delta\theta$, we have
$\frac{\delta\varphi}{\rho_{\sss GP}}= \frac{\delta\rho}{2\rho_{\sss
    GP}} +i\delta\theta$.  Separating real and imaginary parts in
Eq. (\ref{undici}) we obtain \cite{long-paper}
\begin{eqnarray}
  && \nonumber
  \hbar(\partial_t+v_{\sss GP}\partial_x)\delta\theta
  -\frac{\hbar^2}{4m\rho_{\sss GP}}\partial_x
  \left(\rho_{\sss GP}\partial_x \left(\frac{\delta\rho}{\rho_{\sss GP}}\right)\right)
  \\
  \nonumber &&
  +g\delta\rho=-\frac{g\rho_{\sss GP}}{2}g^{(2)}
  -\frac{\hbar}{2}
  (\partial_t + v_{\sss GP}\partial_x)Re\langle \hat\eta\hat\theta\rangle \\
  \label{ventiduea}  && 
  -\frac{\hbar^2}{4m\rho_{\sss GP}}\partial_x
  \left[
    \rho_{\sss GP}\partial_x
    \left(\langle \hat\theta^2\rangle -\frac{\delta(0)}{4\rho_{\sss GP}}
    +\frac{g^{(2)}}{4}\right)
    \right]\ , \\
  \label{ventidueb} &&
  \partial_t\delta\rho +\partial_x\left(v_{\sss GP}\delta\rho
  +\rho_{\sss GP}\delta v +
  Re\langle \rho_{\sss GP}\hat\eta\hat v\rangle\right)=0\ , 
\end{eqnarray}
where $\delta v=\frac{\hbar}{m}\partial_x\delta \theta\ , \hat
v=\frac{\hbar}{m}\partial_x\hat\theta\ , v_{\sss
  GP}=\frac{\hbar}{m}\partial_x \theta_{\sss GP}$.

All expectation values have to be taken on the Gross-Pitaevskii background
$\varphi_{\sss GP}$ using the Bogoliubov-de Gennes equation in the
density-phase representation
\begin{eqnarray}
  \label{ventitrea} &&\hbar(\partial_t+v\partial_x)\hat\theta
  -\frac{\hbar^2}{4m\rho_{\sss GP}}
  \partial_x(\rho_{\sss GP}\partial_x\hat\eta)+g\rho_{\sss GP} \hat\eta=0, \ \
  \\ \label{ventitreb} &&
  (\partial_t+v_{\sss GP}\partial_x)\hat\eta
  +\frac{\hbar}{m\rho_{\sss GP}}
  \partial_x(\rho_{\sss GP}\partial_x\hat\theta)=0\ . \end{eqnarray}
In Eq. (\ref{ventiduea}) 
\begin{equation} \label{ventiquattro}
  g^{(2)}=\langle \hat\eta^2\rangle -\frac{\delta(0)}{\rho_{\sss GP}}\ ,
\end{equation}
which because of the Dirac $\delta(0)$ contribution is ultraviolet
regular in one dimension.

Note that $\langle \hat \theta^2\rangle$ is infrared divergent in one
dimension, but this divergence is killed by acting on it with the
spatial derivative in Eq. (\ref{ventiduea}) \cite{afb2015}.
 This is analogous to the situation in two-dimensional
  gravity, where the two-point function is infrared divergent, but
  contributes to the field equation only through its appearance in
  $\langle \hat T_{\mu\nu}\rangle$, which involves derivatives and is
  therefore infrared regular.
So our one dimensional
backreaction equations (\ref{ventiduea}, \ref{ventidueb}) are both
ultraviolet and infrared finite.

We note that for $v_{\sss GP}=0$ our backreaction equations reduce to
those obtained by Mora and Castin \cite{mc2003} with a more rigorous
approach by discretizing space.  To solve the backreaction equations
one needs to determine the quantum source terms in the whole physical
space. As shown in \cite{mathieu-nicolas} this can be accurately done
only by taking into account also the zero modes solutions of the
Bogoliubov-de Gennes equation and this is particularly critical near
the horizon of a sonic BH. The discussion of this delicate point is
postponed to a future work.

As a simple example of the use of the backreaction equations we
consider the case of a uniform condensate for which there is no
contribution of the zero modes.  For a homogeneous condensate the
field operators can be expanded in terms of plane waves as
\begin{eqnarray}
  \label{venticinquea} \hat\eta= \frac{1}{\sqrt{\rho_{\sss GP}}}
  \int_0^{+\infty}d\omega (u_\omega +v_\omega)\hat b_\omega e^{-i\omega t+ik(\omega)x}
  +h.c., \ \ \ \ \ \ \\
  \label{venticinqueb} \hat\theta=\frac{1}{2i\sqrt{\rho_{\sss GP}}}
  \int_0^{+\infty}d\omega (u_\omega-v_\omega)\hat b_\omega e^{-i\omega t+ik(\omega)x}
  +h.c., \ \ \ \ \ \ \end{eqnarray}
where the dispersion relation reads
\begin{equation} \label{ventisei}
  (\omega -v_{\sss GP}k)^2=c_{\sss GP}^2
  \left(k^2+\xi_{\sss GP}^2\frac{k^4}{4}\right)\ ,
\end{equation}
$c_{\sss GP}=\sqrt{\frac{g\rho_{\sss GP}}{m}}$ is the Gross-Pitaevskii
speed of sound and $\xi_{\sss GP}$($=\hbar(mg\rho_{\sss GP})^{-1/2}$)
the corresponding healing length. The Bogoliubov modes are \cite{mp,
  rpc, mfr}
\begin{equation}\label{ventisettea}
  u_\omega =  \frac{\omega -v_{\sss GP} k+
    \frac{c_{\sss GP}\xi_{\sss GP} k^2}{2}}
  {\sqrt{4\pi  c_{\sss GP}\xi_{\sss GP} k^2\left|
      (\omega-v_{\sss GP}k) \left(\frac{dk}{d\omega}\right)^{-1} \right| }},
\end{equation}
\begin{equation}\label{ventisetteb}
  v_\omega = -\frac{\omega -v_{\sss GP} k-\frac{c_{\sss GP}\xi_{\sss GP} k^2}{2}}
  {\sqrt{4\pi c_{\sss GP}\xi_{\sss GP} k^2\left|
      (\omega-v_{\sss GP}k) \left(\frac{dk}{d\omega}\right)^{-1} \right| }}\ .
\end{equation}

The operators $\hat b_\omega, \hat b_\omega^\dagger$ satisfy standard
Bose commutation rules. The backreaction equations simplify to
($\delta\mu=-\hbar\partial_t\delta\theta$)
\begin{eqnarray} \label{ventottoa}
  &&
g\delta\rho +mv_{\sss GP}\delta v=\delta \mu -\frac{1}{2}g\rho_{\sss
  GP}g^{(2)}\ , \\
\label{ventottob}
&&v_{\sss GP}\delta\rho +\rho_{\sss GP}\delta v
+Re\langle\rho_{\sss GP}\hat\eta\hat v\rangle=const=\delta J\ .
\end{eqnarray}
Using the expansions (\ref{venticinquea},\ref{venticinqueb}) one gets 
\begin{eqnarray}
\label{ventinove} &&g^{(2)}=-\frac{2}{\pi\rho_{\sss GP}\xi_{\sss GP}}\ , \\
\label{trenta} &&Re\langle\rho_{\sss GP}\hat\eta\hat v\rangle =0\ ,
\end{eqnarray}
which we note are independent of $v_{\sss GP}$. So for a condensate at
rest ($v_{\sss GP}=0$) the result reads \cite{mc2003}
\begin{eqnarray}
  \label{trentuno1} \delta\rho &=&
  -\frac{1}{2}\rho_{\sss GP}g^{(2)}=\frac{1}{\pi\xi_{\sss GP}}\ , \\
  \label{trentuno2}
  \delta v &=& 0=\delta \mu\ . 
\end{eqnarray}
From $\mu=g\rho_{\sss GP}$ and $\rho_{\sss GP}=\rho-\delta\rho$ we
obtain the equation of state in the form
\begin{equation}
  \label{trentadue} \mu=g\rho-\frac{g}{\pi\xi}\ .
\end{equation}
Using the thermodynamical relation
$mc^2=\rho(\frac{\partial\mu}{\partial\rho})$ we obtain, for the speed
of sound, \cite{lieb}
\begin{equation} \label{trentaquattro}
     c=\sqrt{\frac{g\rho}{m}}\,
    \sqrt{1-\frac{1}{2\pi\rho\xi}} \ ,
\end{equation}
which corrects  the Gross-Pitaevskii result $c_{\sss
  GP}$.  Galilean invariance imposes that this result
  remains unaltered in the presence of a finite flow velocity
$v_{\sss GP}\neq 0$. 
  Eqs. \eqref{trentadue} and \eqref{trentaquattro} indicate
  that the small parameter of our approach is the quantity
  $1/\rho\xi$, which is of order $2\times 10^{-2}$ in Steinhauer's
  experiments \cite{jeff2016}.

From Eq. (\ref{trentaquattro}) we can calculate the correction to the
speed of sound and hence the correction to Gross-Pitaevskii Mach
number $M_{\sss GP}=\frac{v_{\sss GP}}{c_{\sss GP}}$, namely
\begin{equation} \label{trentacinque} 
\frac{\delta M}{M_{\sss GP}}
=\frac{\delta v}{v_{\sss GP}}-\frac{\delta c}{c_{\sss GP}}
=-\frac{1}{2}\frac{\delta\rho}{\rho_{\sss GP}}
+\frac{1}{4\pi\xi_{\sss GP}\rho_{\sss GP}}
=-\frac{1}{4\pi\xi_{\sss GP}\rho_{\sss GP}}\ .
\end{equation}
Here the backreaction, by changing the density and hence the speed of
sound, decreases the Mach number from the Gross-Pitaevskii value.

Coming to non homogeneous configurations, most of the sonic BH flow
profiles discussed in the literature and also experimentally realized
contain a subsonic asymptotic region (say $x\to +\infty$) where the
condensate is homogeneous and, similarly, a homogeneous supersonic
asymptotic region (say $x\to -\infty$ for a fluid flowing from right
to left).  Let us call the first asymptotic region $u$ (for upstream)
and the other $d$ (downstream).  In these regions the backreaction
equations for a stationary asymptotically homogeneous flow read
\begin{eqnarray} \label{trentunoa}
  && g\delta\rho_\alpha +mv_{\sss
  GP}^\alpha\delta v_\alpha=\delta\mu-\frac{1}{2}g\rho_{\sss
    GP}^\alpha g_\alpha^{(2)}\ , \\
  \label{trentunob} && v_{\sss
  GP}^\alpha \delta\rho_\alpha +\rho_{\sss GP}^\alpha \delta
  v_\alpha=\delta J-J_\alpha\ ,
\end{eqnarray}
where $\alpha=u,d$ and
$J_\alpha=Re\langle \rho_{\sss GP}\hat\eta\hat v\rangle_\alpha$. The
quantities labeled with $\alpha$ {\color{red} in}
these equations are asymptotic
quantities in the far upstream and downstream regions.  The general
structure of $g_\alpha^{(2)}$ for a sonic BEC BH is the following
\begin{equation} \label{trentatrea}
g_\alpha^{(2)}=-\frac{2}{\pi\rho_{\sss GP}^\alpha\xi_{\sss GP}^\alpha}
+g_\alpha^{(H)}\ ,
\end{equation}
where
the first term reproduces Eq. (\ref{ventinove}) while $g_\alpha^{(H)}$
is induced by Hawking radiation and vanishes in the absence of a sonic
horizon. This splitting is reminiscent of the analogous one present in
the 2D gravitational $\langle T_{\mu\nu}\rangle$ in the Unruh state,
which can be splitted in a vacuum polarization (Boulware) term plus
the Hawking radiation contribution. On the other hand, the flux term
$J_\alpha$ comes (like in gravity) only from Hawking radiation.

Simple manipulations of eqs. (\ref{trentunoa}, \ref{trentunob}) lead
to the asymptotic variations of the condensate density as
\begin{equation} \label{quarantuno}
\delta \rho_\alpha = \frac{\delta\mu -
  \frac{1}{2}g\rho_{\sss GP}^\alpha
  g_\alpha^{(2)}
  -\frac{mv^\alpha_{\sss GP}}{\rho_{\sss GP}^\alpha}
  (\delta J - J_\alpha)}
       {g\left(1-\frac{(v_{\sss GP}^\alpha)^2}{(c_{\sss GP}^\alpha)^2}\right)}\ ,
\end{equation}
and
$\delta v_\alpha$ can be obtained inserting this in the continuity
equations (\ref{trentunob}).  For our perturbative approach to be
valid (i.e. $\frac{\delta \rho_\alpha}{\rho_{\sss GP}^\alpha}\ll 1$)
even in the case when the asymptotic regions approach the sonic limit
(i.e. $|v_{\sss GP}^\alpha|\to c_{\sss GP}^\alpha$) $\delta\mu$ and
$\delta J$ have to be such that the numerator of
Eq. (\ref{quarantuno}) vanishes in this limit. Example of interesting
sonic BH profiles
\cite{lrcp} with the explicit calculations of the source terms
$g^{(2)}$ and $J_\alpha$ and the corresponding $\delta\rho_\alpha$ and
$\delta v_\alpha$ are given elsewhere \cite{long-paper}.

As we have said the calculation of the quantum source terms in the
backreaction equations for BEC black holes is much more complicated
than in gravity because of the fourth order dispersion relation and
the contribution of zero modes (which are not present in gravity) in
particular in the horizon region. However we have now a consistent
theoretical scheme to attack the backreaction problem for one
dimensional condensates and moreover we should not forget that in the
BEC case we have a significant advantage we do not have in gravity: we
can have experimental insights to guide our theoretical
investigations.

\textbf{Acknowledgments:} A.F. acknowledges partial financial support
by the Spanish Grant PID2023-149560NB-C21 funded by
MCIN/AEI/10.13039/501100011033 and FEDER, European Union, and the
Severo Ochoa Excellence Grant CEX2023-001292-S. G. C. acknowledges
partial financial support by the Deutsche Forschungsgemeinschaft
funded Research Training Group "Dynamics of Controlled Atomic and
Molecular Systems" (Grant No. RTG 2717).

\end{document}